\documentclass{llncs}

\usepackage{algorithm}
\usepackage{algorithmic}
\usepackage{amsfonts}
\usepackage{amsmath}
\usepackage{amssymb}
\usepackage{bm}
\usepackage[justification=centering]{caption}
\usepackage{cite}
\usepackage{color}
\usepackage{graphicx}
\usepackage{epstopdf}

\begin{document}

\pagestyle{headings}

\renewcommand{\algorithmicrequire}{\textbf{Input:}}
\renewcommand{\algorithmicensure}{\textbf{Output:}}
\renewcommand{\algorithmicprint}{\textbf{break}}

\title{A Faster Algorithm to Build New Users Similarity List in Neighbourhood-based Collaborative Filtering}

\author{Zhigang Lu \and Hong Shen}

\institute{School of Computer Science,\\The University of Adelaide, SA 5005, Australia}

\maketitle

\begin{abstract}
Neighbourhood-based Collaborative Filtering (CF) has been applied in the industry for several decades, because of the easy implementation and high recommendation accuracy. As the core of neighbourhood-based CF, the task of dynamically maintaining users' similarity list is challenged by cold-start problem and scalability problem. Recently, several methods are presented on solving the two problems. However, these methods applied an $O(n^2)$ algorithm to compute the similarity list in a special case, where the new users, with enough recommendation data, have the same rating list. To address the problem of large computational cost caused by the special case, we design a faster ($O(\frac{1}{125}n^2)$) algorithm, TwinSearch Algorithm, to avoid computing and sorting the similarity list for the new users repeatedly to save the computational resources. Both theoretical and experimental results show that the TwinSearch Algorithm achieves better running time than the traditional method.
\end{abstract}

\section{Introduction}
\label{sec:introduction}
In recommender systems, Collaborative Filtering (CF) is a famous technology with three main popular algorithms\cite{LIU2011}, i.e., neighbourhood-based methods\cite{EKSTRAND2011}, association rules based prediction\cite{SARWAR2001}, and matrix factorisation\cite{KOREN2009}. Among these algorithms, neighbourhood-based methods are the most widely used in the industry because of its easy implementation and high prediction accuracy.

The core of neighbourhood-based CF methods is the computation of a sorted similarity list for every user. The task of dynamically maintaining a similarity list is important in a neighbourhood-based recommender system, as the creation of new users and the rate updates of old users will result in an update of the similarity list frequently. Accordingly, there are two main research problems in recommender systems, one is the cold-start problem, the other is the scalability problem \cite{PAGARE2013}. 
Recent research\cite{BOBADILLA2012, LIU2014, ADAMOPOULOS2014, LIKA2014} on addressing cold-start problem focus on improving the prediction accuracy with the limit rates information. While, 
Some of the solutions\cite{PAPAGELIS2005, LIU2010, YANG2012, HUANG2015} to the scalability problem work on decreasing the computational cost by linking the new similarity list with the old one.

Different from the two classic problems, we notice a special case where the methods listed above do not work well. In this case, we assume the new users have enough rating data to build reliable similarity list, also, they have totally the same ratings list. The methods aim to solve cold-start problem or scalability problem will treat this special case as a normal input of recommender systems, then apply an $O(n^2)$ algorithm to compute the new users' similarity list. Considering the number of users in a recommender system, $n$, is usually very large, the computational cost of the above method will be very large.

To address the problem of large computational cost caused by the special case, we design a faster ($O(\frac{1}{125}n^2)$) algorithm, TwinSearch Algorithm, to avoid computing and sorting the similarity list for the new users repeatedly to save the computational resources. Moreover, we compare the running time of TwinSearch algorithm and the traditional similarity computation method in two real-world data sets on both user-based and item-based CF. Both theoretical and experimental results show that the TwinSearch algorithm achieve better running time than the traditional method. To the best of our knowledge, we are the first to consider this special case in recommender systems.

The rest of this paper is organised as follows: Firstly, in Section \ref{sec:RW}, we discuss the existing technologies to dynamically maintain the similarity list in recommender systems. Afterwards, in Section \ref{sec:ALG}, we present a faster algorithm to build new users similarity list in neighbourhood-based CF recommender systems, and analyse the time complexity of our novel algorithm theoretically. Next, in Section \ref{sec:EE}, the experimental analysis of our algorithm on the performance of running time are provided. Finally, we conclude with a summary in Section \ref{sec:CON}.

\section{Related Work}
\label{sec:RW}
In this section, we briefly summarise some of the current works on addressing cold-start problem and scalability problem.

Some of the methods on cold-start problem focus on the improvement of prediction accuracy, due to the lack of enough rating data of new users. Theses methods gain better prediction performance by applying different strategy. For example, Bobadilla et al.\cite{BOBADILLA2012} presented a new similarity measure with optimization based on neural learning, which shows the much better results than current metrics, such as cosine similarity measure. Liu et al.\cite{LIU2014} showed an interesting phenomenon that to link a cold-start item to inactive users will give this new item more chance to appear in other users’ recommendation lists. Adamopoulos et al.\cite{ADAMOPOULOS2014} applied probabilistic method to select the $k$ neighbours from the entire candidate list, rather the $k$ nearest candidate, to avoid the low prediction accuracy due to the lack of rates data. Lika et al.\cite{LIKA2014} proposed an approach which incorporates classification methods in a pure CF system while the use of demographic data help for the identification of other users with similar behaviour.

Some of the solutions to scalability problem proposed the methods based on incremental updates of user-to-user and item-to-item similarity. These methods achieve faster and high-quality recommendation than the traditional CF. Papagelis et al.\cite{PAPAGELIS2005} proposed an incremental method which quickly updates user's similarity list when the user adds/rates new items in the recommender systems. Liu et al.\cite{LIU2010} presented the temporal relevance measure for ratings at different time steps and developed online evolutionary collaborative filtering algorithms by introducing this measure into $k$NN algorithms and incrementally computing neighbourhood similarities, which achieve both better time and space complexity. Inspired by \cite{PAPAGELIS2005}, Yang et al.\cite{YANG2012} developed the user-based incremental similarity update method to an corresponding item-based method. Huang et al.\cite{HUANG2015} proposed a practical item-based CF algorithm on big data environment, with the super characteristics such as robust to the implicit feedback problem, scalable incremental update and real-time pruning.

Unfortunately, the current methods on cold-start problem and scalability problem do not work well on a special case: the new users, with enough recommendation data, have the same rating list ($k$ Nearest Neighbour ($k$NN) attack\cite{CALANDRINO2011} can be taken as an example of our special case, which creates $k$ same fake users with at least 8 rated items into the recommender system). The reasons are the solutions to cold-start problem only work on the new users which have not been gathered sufficient information, and the methods concentrating on scalability problem only work on the old users who have already have a similarity list. Naturally, when facing the special case, the above methods have to apply the traditional similarity computation method which yields in $O(n^2)$ time complexity. Considering the number of users in a recommender system, $n$, is usually very large, the computational cost of the above method will be very large. Therefore, it is necessary to gain a faster algorithm to build the new users similarity list in our special case.

\section{The TwinSearch Algorithm}
\label{sec:ALG}

\subsection{Algorithm Design}
In this section, we define the users who have the same rating list as $twin$ $users$. To address the large computational cost due to the special case: the new users, with enough recommendation data, have the same rating list, we aim to avoid computing and sorting the similarity list for the new users repeatedly to save the computational resources. Since the new users are the same, our strategy to avoid repeated computation is searching the twin user from the system, then copying the twin user's similarity list to the new user directly.

According to the properties of the similarity in recommender systems, we know that if two users are twin user, i.e., $u_a = u_b$, then the similarity between an arbitrary user $u_i$ and $u_a$, $u_b$ are equal, i.e., $sim_{ai} = sim_{bi}$. Based on the definition of twin user, the ratings on any item $i$ of twin user are equal, i.e., $r_{ai} = r_{bi}$. Therefore, we have the following relationships:
\begin{equation}
\label{eq:r1}
u_{a}=u_{b} \Rightarrow sim_{ai}=sim_{bi}
\end{equation}
\begin{equation}
\label{eq:r2}
u_{a}=u_{b} \Leftrightarrow r_{ai}=r_{bi}
\end{equation}
Relationship \ref{eq:r1} helps us to find the potential twin users from the system, Relationship \ref{eq:r2} helps us to find the exact twin user from the potential ones. Now we design the TwinSearch Algorithm to find and copy the twin user's similarity list to the new user by relationship \ref{eq:r1} and \ref{eq:r2}.

\begin{algorithm}[!ht]
\caption{TwinSearch Algorithm.}
\label{algo:serialAlgorithm}
\begin{algorithmic}[1]
\REQUIRE ~~\\
A user-item rating set, $\mathcal{R}$, with $n$ users and $m$ items; a user-user sorted similarity matrix, $\mathcal{S}$; a new user, $u_0$, with several ratings on different items; a constant, $c \in \mathbb{Z}^+$.
\ENSURE ~~\\
The new user $u_0$'s similarity list.
\STATE Select $c$ random users, $u_i^*$, $i \in [1,c]$;
\FOR {$i=1$ to $c$}
\STATE compute similarity between user $u_0$ and $u_i^*$, $sim_{0i}$;
\STATE search $u_i^*$'s similarity list $\mathcal{S}_i$ for a $Set_i=\{u_x | sim_{ix}=sim_{0i}\}$;
\IF{$sim_{0i}=1$}
\STATE add $u_i^*$ to $Set_i$;
\ENDIF
\ENDFOR
\STATE Compute the intersection $Set_0$ of the $c$ $Set_i$s, $Set_0=\bigcap_{i=1}^c Set_i$;
\FOR{$i=1$ to $|Set_0|$}
\IF{$r_{ij} = r_{0j}$, $j \in [1,m]$}
\STATE copy the similarity list of $u_i \in Set_0$ to $u_0$;
\STATE $break$; 
\ENDIF
\ENDFOR
\RETURN The new user $u_0$'s similarity list.
\end{algorithmic}
\end{algorithm}
In line 4, we search the potential twin users by Relationship \ref{eq:r1}. In line 9, we narrow the size of the final potential twin user set $Set_0$ by intersecting the $c$ bigger potential twin user set $Set_i$. The for loop in lines 10-15 find the twin user from the potential twin users' set by Relationship \ref{eq:r2}. Our algorithm can be worked in both user-based and item-based CF, in this section, we present the TwinSearch algorithm from the perspective of the user-based methods, and this can be applied to item-based methods in a straightforward way.

\subsection{Time Complexity Analysis}
We select the $c$ random users in line 1 in $O(c)$. The for loop in lines 2-8 contributes $O(c(m+\log{n}))$ to running time, if we use binary search in line 4. To compute the intersection $Set_0$ in line 9, it takes $O(cn)$ time. The for loop in lines 10-15 requires $O(|Set_0|m)$ time, if we use the link list as the similarity matrix $\mathcal{S}$ data structure. Therefore, the total running time of Algorithm \ref{algo:serialAlgorithm} is $O(|Set_0|m + c(m+\log{n}))$.

Now we focus on the value of $|Set_0|$. Because $Set_0=\bigcap_{i=1}^c{Set_i}$, $|Set_0| \leq \min\{|Set_i|\}$, i.e., $|Set_0| = \max\{\min\{|Set_i|\}\}$, $i \in [1, c]$. As the values in a specific $Set_i$ are equal, $Set_i$ must be included in one sub-list of the original similarity list. The sub-list is produced by partitioning the similarity list with the similarity value. For example, suppose that we have $x$ sub-lists, then the similarity value in each sub-list is in the range of $[0, \frac{1}{x}), [\frac{1}{x}, \frac{2}{x}), \cdots, [1-\frac{1}{x}, 1.0]$ correspondingly. Thus, the upper bound of $|Set_0|$ must be less than the size of largest sub-list.

\begin{figure}[!htbp]
\centering
\includegraphics[width=0.7\textwidth]{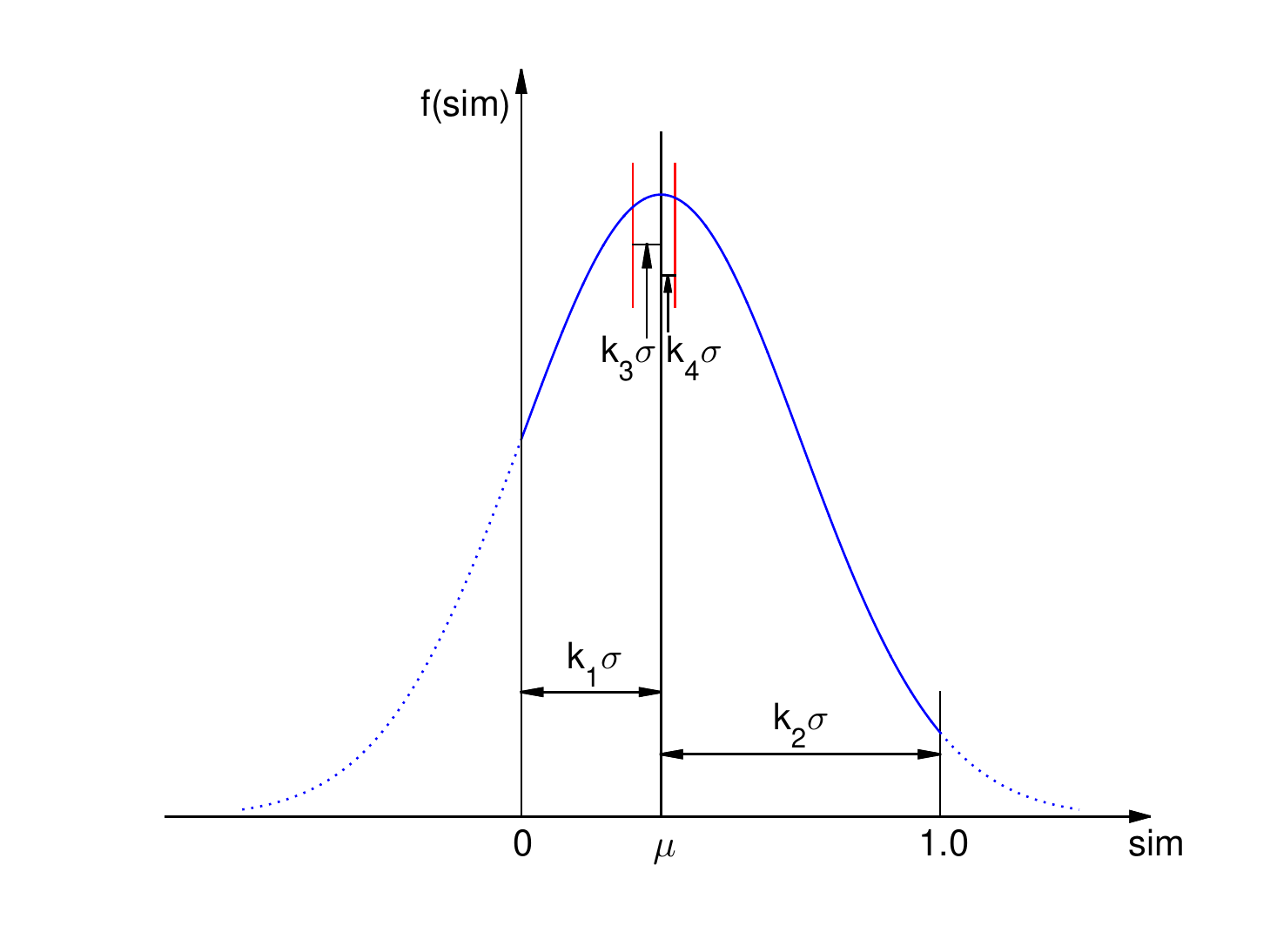}
\caption{Distribution of user's similarity list}
\label{fig:simDis}
\end{figure}

Moreover, Wei et al.\cite{WEI2005} showed that any user's similarity list obeys a specific Gaussian distribution in recommender systems. In this paper, because of the value of similarity, we set the sample range in [0, 1.0].
Since for any Gaussian distributions, more than 99.99\% samples are in the range of $[\mu-4\sigma, \mu+4\sigma]$, we fix the similarity value range [0, 1.0] within $\mu \pm 4\sigma$ in this paper. Figure \ref{fig:simDis} shows the basic statistic settings of one similarity list, where 
the greatest size sub-list's similarity value range is between $[\mu - k_3\sigma, \mu + k_4\sigma]$. Therefore, we have the size of the sub-list with the most number of users, $s = \frac{\text{Area under the Gaussian distribution curve between }\mu-k_3\sigma \text{ and } \mu+k_4\sigma}{\text{Area under the Gaussian distribution curve between 0 and 1.0}} \times n$. According to the property of Gaussian distribution, we rewrite the expression of $s$ as:
\begin{equation}
\label{eq:set0}
\begin{array}{ccl}
s & = & \frac{\Phi(\frac{\mu+k_4\sigma-\mu}{\sigma})-\Phi(\frac{\mu-k_3\sigma-\mu}{\sigma})}{\Phi(\frac{\mu+k_2\sigma-\mu}{\sigma})-\Phi(\frac{\mu-k_1\sigma-\mu}{\sigma})} \times n\\
 & = & \frac{\Phi(k_3)+\Phi(k_4)-1}{\Phi(k_1)+\Phi(k_2)-1} \times n,
\end{array}
\end{equation}
where $\Phi(x)$ is the cumulative distribution function of standard Gaussian distribution. Our goal is to find the maximum value of $s$.

In fact, for a specific Gaussian distribution and a partition parameter $x$, the area under the Gaussian distribution curve between $\mu-k_3\sigma$ and $\mu+k_4\sigma$ is fixed. But, when $k_1 = k_3$, the area under the Gaussian distribution curve between 0 and 1.0 reaches the minimum value. Thus, when $k_1 = k_3$, the value of $s$ is maximum. Then, we have the following linear programming:
\begin{equation}
\label{eq:lp}
\begin{array}{ll}
\text{maximise} & s=\frac{\Phi(k_3) + \Phi(k_4) - 1}{\Phi(k_1) + \Phi(k_2) - 1} \times n\\
\text{subject to} & \mu - k_1\sigma = 0\\
 & \mu + k_2\sigma = 1\\
 & \mu - k_3\sigma = 0\\
 & \mu + k_4\sigma = \frac{1}{x}\\
 & 0 \leq k_1 \leq 4, 0 < k_2 \leq 4\\
 & 0 \leq k_3, 0 < k_4.
\end{array}
\end{equation}

According to the properties of the cumulative distribution function of standard Gaussian distribution, we have the solution for linear programming \eqref{eq:lp}: $k_1 = 0$, $k_2 = 4$, $k_3 = 0$, $k_4 = 0.01$. Then we have the maximum $s = \frac{1}{125}n$ which is the upper bound of $|Set_0|$. In this paper, we assume $c \ll \frac{1}{125}n$. Therefore, the overall running time of TwinSearch Algorithm \ref{algo:serialAlgorithm} is $O(\frac{1}{125}mn)$, which is much less than the running time ($O(mn)$) of traditional similarity computation method. In this paper, we assume there are $k$ new same users will be created in the system, so the total running to build the $k$ users in traditional similarity computation method is $O(kmn)$, while in the TwinSearch algorithm, it is $O((1+\frac{k-1}{125})mn)$.


%
%
%
%
%
%

\section{Experimental Evaluation}
\label{sec:EE}
In this section, we use the real-world datasets to evaluate the performance on time complexity of TwinSearch Algorithm and traditional similarity computation method. We begin by the description of the datasets, then perform a comparative analysis of our algorithm and the traditional similarity computation.

\subsection{Datasets and Experimental Settings}
In the experiments, we use two real-world datasets, MovieLens dataset\footnote{http://www.grouplens.org/datasets/movielens/} and Douban\footnote{http://www.douban.com} (one of the largest rating websites in China) film dataset\footnote{https://www.cse.cuhk.edu.hk/irwin.king/pub/data/douban}. The MovieLens dataset consists of 100,000 ratings (1-5 integral stars) from 943 users on 1682 films, where each user has rated at least 20 films, each film has been rated by 20$-$250 users. The Douban film dataset contains 16,830,839 ratings (1-5 integral starts) from 129,490 unique users on 58,541 unique films \cite{MA2011}. All the experiments are implemented in MATLAB 8.5 (64-bit) environment on a PC with Intel Core2 Quad Q8400 processor (2.67 GHz) with 8 GB DDR2 RAM.

\subsection{Experimental Results}
\begin{figure}
\begin{minipage}[!h]{0.5\linewidth}
\centering
\includegraphics[width=2.42in]{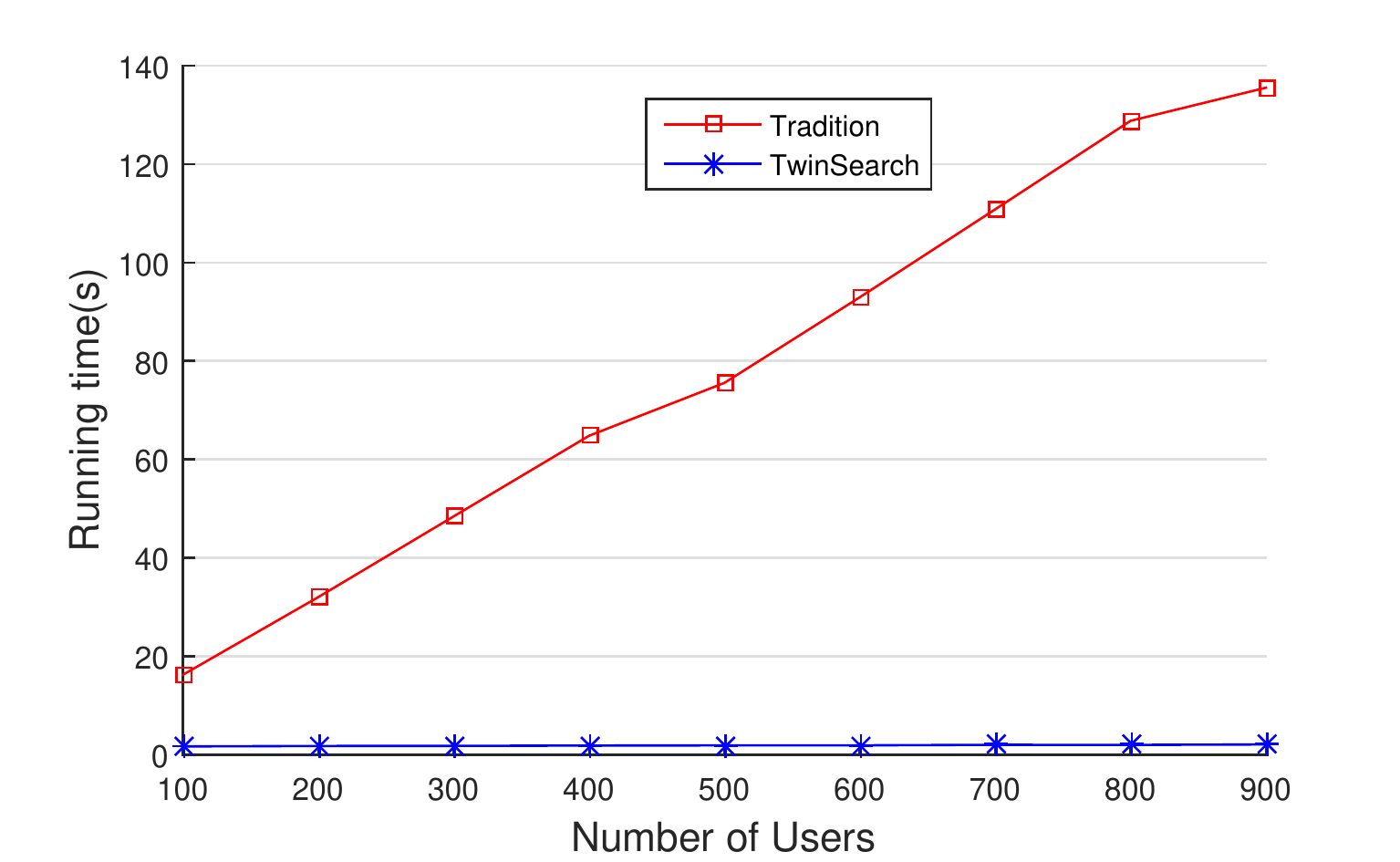}
\caption{Running time of User-based CF on MovieLens}
\label{fig:userM}
\end{minipage}
\begin{minipage}[!h]{0.5\linewidth}
\centering
\includegraphics[width=2.42in]{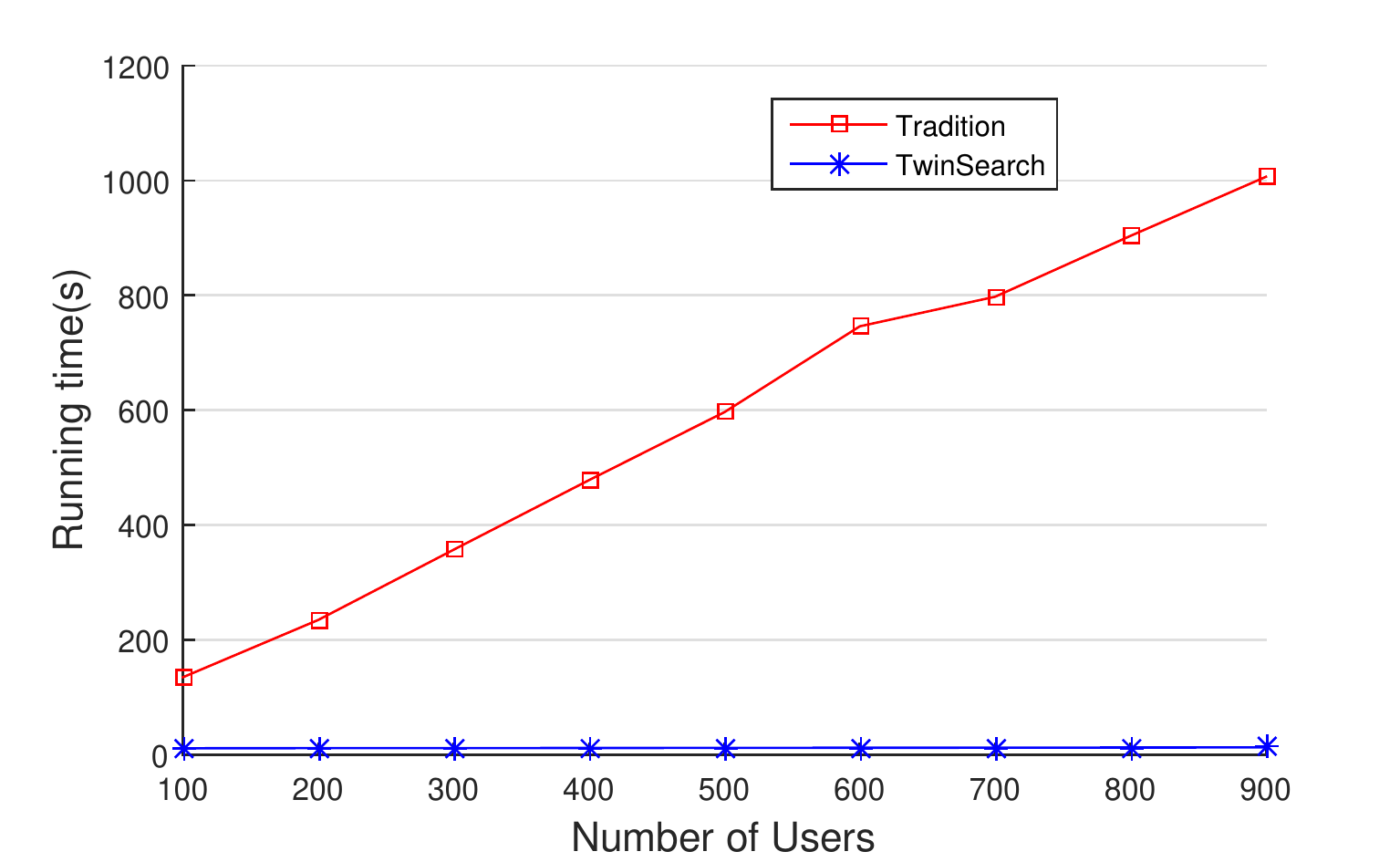}
\caption{Running time of User-based CF on Douban film}
\label{fig:userD}
\end{minipage}
\end{figure}

\begin{figure}
\begin{minipage}[!h]{0.5\linewidth}
\centering
\includegraphics[width=2.42in]{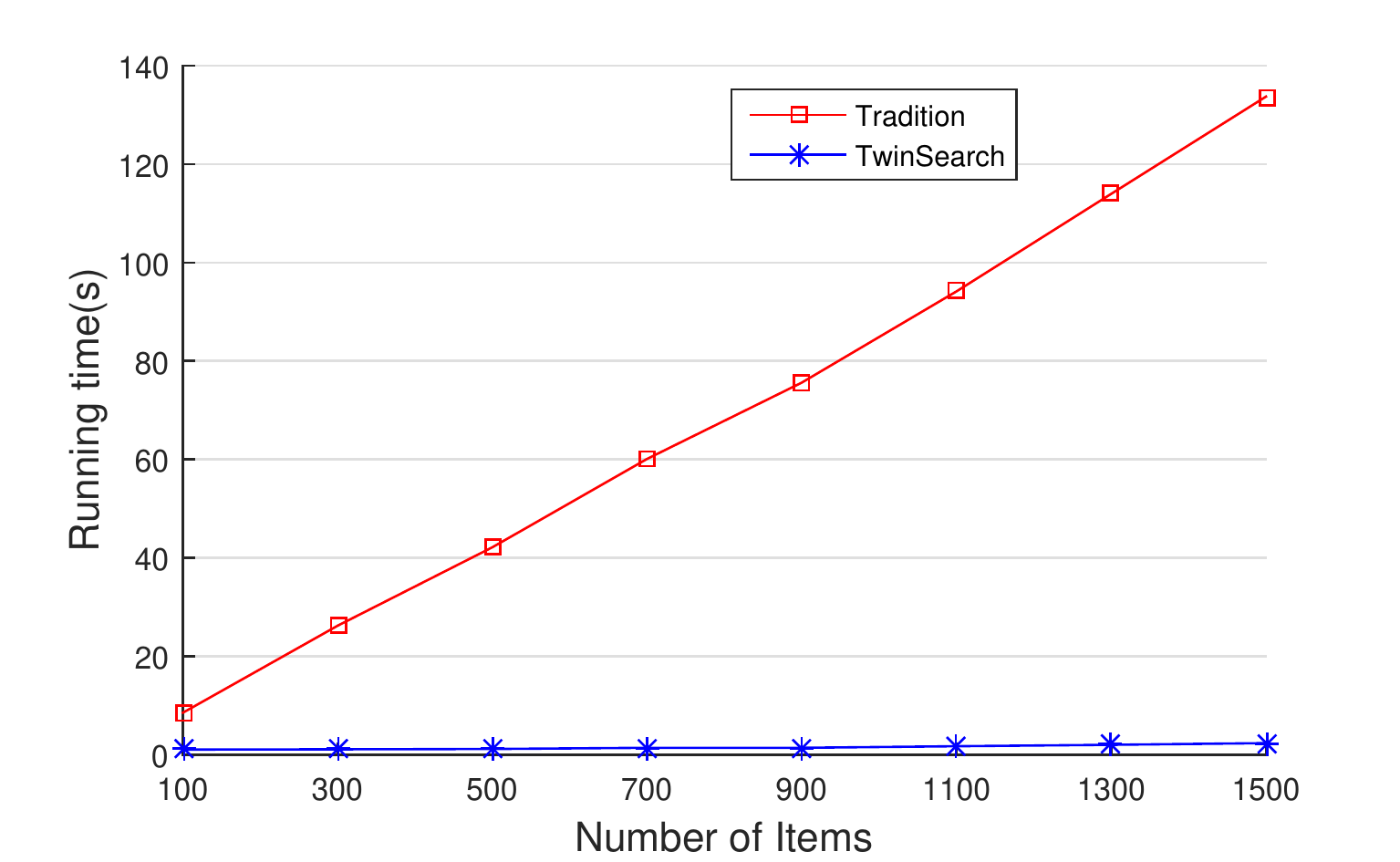}
\caption{Running time of Item-based CF on MovieLens}
\label{fig:itemM}
\end{minipage}
\begin{minipage}[!h]{0.5\linewidth}
\centering
\includegraphics[width=2.42in]{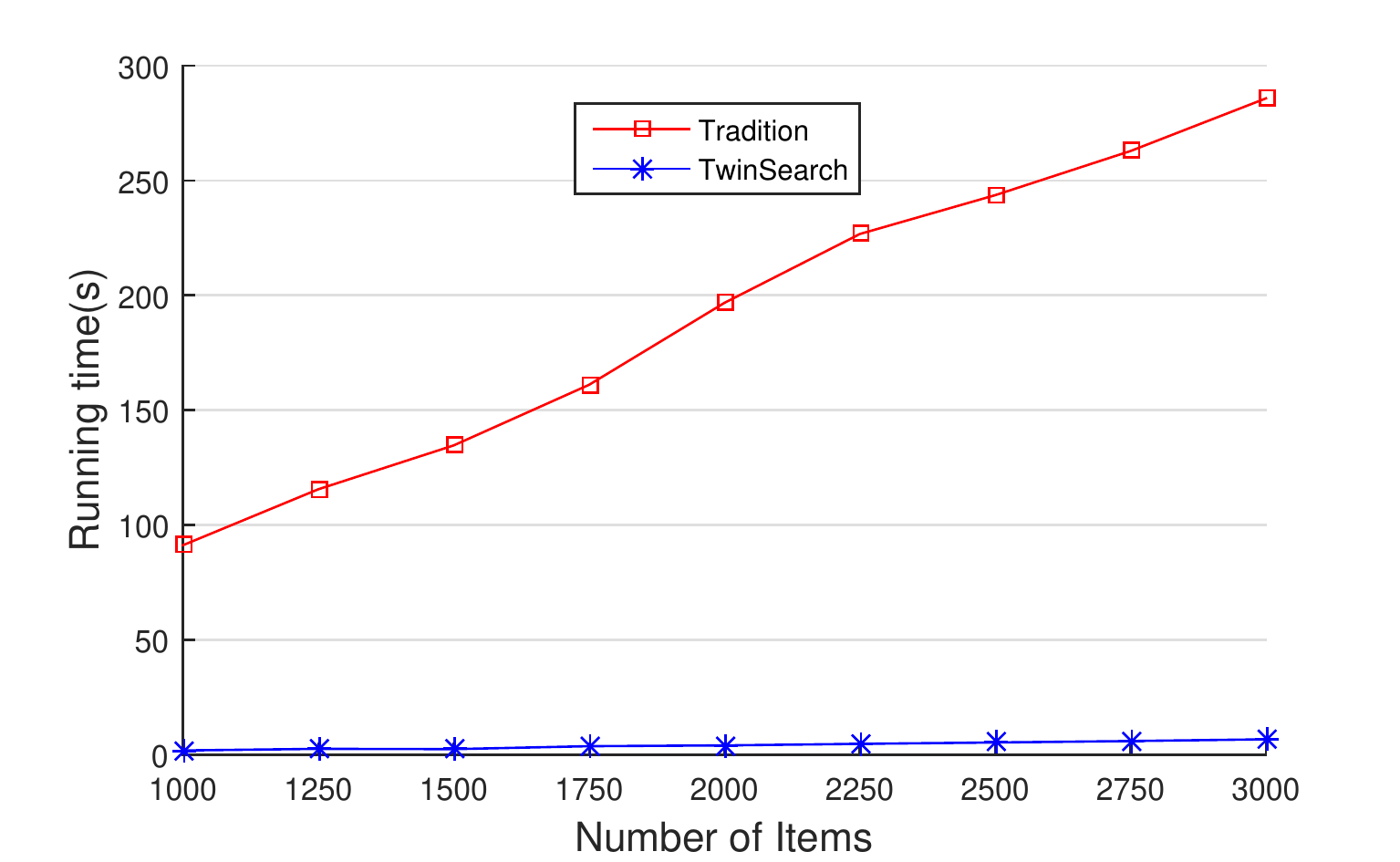}
\caption{Running time of Item-based CF on Douban film}
\label{fig:itemD}
\end{minipage}
\end{figure}

We design 4 experiments (Figure \ref{fig:userM} to \ref{fig:itemD}) to evaluate the running time for $k$ new user with same ratings on the above two data sets in both user-based and item-based CF. We use cosine similarity metric as the traditional similarity computation method, and set $k = 30$ in the 4 experiments. From the 4 figures, we can see that the TwinSearch algorithm achieves much better performance on time complexity than the traditional similarity computation method.

\vspace{-4em}

\section{Conclusion}
\label{sec:CON}
Neighbourhood-based Collaborative Filtering (CF) play an important role in e-commerce, because of the easy implementation and high recommendation accuracy. Two classic problems, cold-start problem and scalability problem, challenge the task of dynamically maintaining similarity list in neighbourhood-based CF. Recently, several methods are presented on solving the two problems. However, these methods applied a traditional $O(n^2)$ algorithm to compute the similarity list in a special case: the new users, with enough recommendation data, have the same rating list. To address the problem of large computational cost due to the special case, we design a faster ($O(\frac{1}{125}n^2)$) algorithm to build new users' similarity list, which avoids computing and sorting the similarity list to save the computational resources. Both theoretical and experimental results show that our algorithm achieves better running time than the traditional method.

\bibliographystyle{splncs}
\bibliography{zhigang}

\clearpage
\end{document}